# A compact frequency-stabilized pump laser for wavelength conversion in long-distance quantum communication

Kohei Ikeda,[1] Yusuke Hisai,[1] Kazumichi Yoshii,[1] Hideo Kosaka,[1] Feng-Lei Hong[1] Tomoyuki Horikiri,[1,*]

[1]Department of Physics, Graduate School of Engineering Science, Yokohama National University, Yokohama 240-8501, Japan.
*Corresponding author: horikiri-tomoyuki-bh@ynu.ac.jp

We demonstrate a compact frequency-stabilized laser at 1064 nm using Doppler-free saturation absorption spectroscopy of molecular iodine. The achieved laser frequency stability and linewidth are $5.7 \times 10^{-12}$ (corresponding to an uncertainty of the laser frequency of 1.6 kHz) and 400 kHz, respectively. The developed frequency-stabilized laser can be used as a pump laser for wavelength conversion from visible to telecom (or vice versa) to connect quantum memories utilizing nitrogen-vacancy centers in diamond at remote nodes in fiber-based quantum communication.

http://dx.doi.org/10.1364/AO.99.099999

## 1. INTRODUCTION

Quantum communication has been expected to realize unconditionally secure communications. For long-distance quantum communication, the development of quantum repeaters is necessary [1–3]. In quantum repeater nodes, quantum memories are used to transfer the quantum states of propagating photons to electron or nuclear spin states of the memory material. This process enables the entanglement between distant quantum memories. The wavelengths of the photons absorbed and emitted by the quantum memories are generally located in the visible range, for example, $Pr^{3+}:Y_2SiO_5$ (606 nm) [4], nitrogen-vacancy (NV) centers in diamond (637.2 nm) [5] and silicon-vacancy centers in diamond (738 nm) [6]. However, photons of telecommunication wavelengths (1.5 μm) are more suitable for low-loss propagation in optical fibers. Therefore, to realize long-distance quantum communication based on quantum repeaters, visible (or near infrared)-to-telecommunication wavelength conversion is inevitable (see also Fig. 1), which has been intensively studied for various quantum memories [7-10]. For wavelength conversion, a pump laser is used to fulfill energy conservation. For efficient coupling with the narrow linewidth of quantum memories, frequency stabilization of the pump laser is necessary for long-term stable operation. In the case of a quantum memory based on NV centers in diamond–which is one of the most promising quantum memories–the wavelength of the absorbed/emitted photons is 637.2 nm with a linewidth of around 10 MHz [11]. Hence, the pump lasers used for wavelength conversion on both sides of the optical fiber must be frequency-stabilized.

Laser frequency stabilization is usually realized by using i) stable optical cavities, or ii) atomic and molecular resonances as frequency references. In the former case, laser frequency stabilization can be performed at an arbitrary wavelength within the coating specification of the cavity, but may need a high-precision wavemeter to determine the longitudinal mode of the cavity. In addition, frequency drift of the optical cavity may limit the applicability in some cases. In the latter case, the laser frequency is determined based on the absolute frequency of the atomic or molecular resonance. However, in this case, laser frequency stabilization can only be realized at the fixed frequencies of the atomic and molecular resonances.

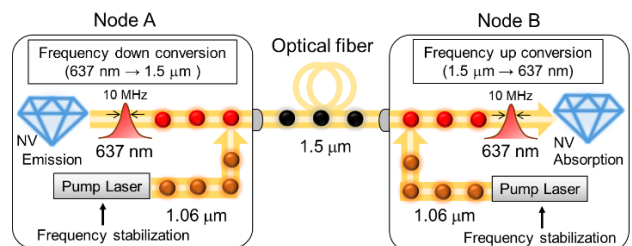

Fig. 1. Schematics of fiber-based quantum communication. Nitrogen-vacancy (NV) centers in diamond are adopted as quantum memories. In this figure, photons from node A are sent to node B.

In this paper, we demonstrate the frequency stabilization of a compact pump laser at 1064 nm for wavelength conversion between 637.2 nm and telecom wavelengths. Laser frequency stabilization is achieved by using Doppler-free spectroscopy of molecular iodine, aiming to control the linewidth and keep the drift of the laser frequency under 10 MHz. The pump laser used in the present experiment is compact, produces a narrow linewidth, and emits high power–in hundreds of mW range. Therefore, this laser meets the experimental requirements for wavelength conversion using NV centers in diamond. However, the commercially-available laser contains no piezoelectric elements for frequency control. We proposed and demonstrated the usage of the driving current of the pump diode in the solid-state laser as an input port for frequency control. We successfully locked the laser frequency without any external frequency modulators, such as an acousto-optic modulator (AOM). This has resulted in the construction of a compact and reliable pump laser system, to be loaded in the quantum repeater nodes.

## 2. EXPERIMENT

Figure 2 shows a schematic diagram of the compact iodine-stabilized pump laser system. A commercially available laser (Showa Opt. JUNO) was used as the light source in the system. This laser contains a laser controller and a laser head, in which a laser diode (LD) pumps the solid-state (Nd:YVO4) laser. The pumped Nd:YVO4 laser oscillates at 1064 nm. The laser cavity consists of a high-reflection-coated Nd:YVO4 crystal surface and a concave mirror to reduce its linewidth (< 1 MHz). The total cavity length (including the 1 mm crystal thickness) is around 20 mm. The dimensions of the laser head are 95 mm × 28 mm × 37 mm (length × width × thickness). The temperature of the compact laser is stabilized by a peltier cooler attached to the bottom of the laser head. Originally, this laser was not designed for frequency stabilization and has no specified frequency tuning knobs. However, the pump laser power changes not only the output power but also the frequency of the Nd:YVO4 laser. The variation of the pump laser power changes the Nd:YVO4 laser temperature, which results in a variation of the refractive index of the laser crystal and consequently the cavity optical path length. Therefore, there is a possibility of frequency control and stabilization via the pump laser power [12, 13].

A 2-cm-long periodically poled KTiOPO4 (PPKTP) crystal was used for second harmonic generation (SHG) at 532 nm. The SHG light was separated from the fundamental light by a dichroic mirror. The fundamental light can be used for wavelength conversion for long-distance quantum communication and/or laser frequency measurement. On the other hand, the 532 nm beam was sent to a saturated absorption spectrometer with a 40-cm-long iodine cell. The spectroscopy of molecular iodine was performed based on the third harmonic technique [14, 15] by applying frequency modulation on the Nd:YVO4 laser through the pump laser power. The laser beam transmitted through the iodine cell (pump beam) was reflected to the cell by a plane mirror. The reflected beam (probe beam) was overlapped with the pump beam. A quarter-wave plate ($\lambda/4$) was set between the iodine cell and the plane mirror, which rotated the polarization of the probe beam by 90°. Consequently, the probe beam was reflected by a polarization beam splitter (PBS) and sent to a photodetector (PD).

The Nd:YVO4 laser frequency was modulated by adding a sinusoidal signal of 2.6 kHz to the injection current of the pump LD. On the other hand, the obtained saturation absorption signal was demodulated at 7.8 kHz. Usually, the third-harmonic technique is operated with a relatively high modulation frequency—for example hundreds of kHz—to obtain a low-noise signal. We set the modulation frequency at a relatively low frequency because we had to add the modulation to the pump LD through the laser controller which has a relatively slow response. The demodulated third derivative signal was fed back to the injection current of the pump LD through a servo controller and the laser controller. The waveform of the third derivative saturation absorption signal was also recorded using a data acquisition device (DAQ). To evaluate the stability of the laser frequency, we measured the heterodyne beat frequency between the compact laser and an iodine-stabilized Nd:YAG laser [16] using the fundamental light. The beat signal was observed by a spectrum analyzer and measured by a frequency counter. The absolute frequency of the iodine-stabilized Nd:YAG laser was measured using an optical frequency comb.

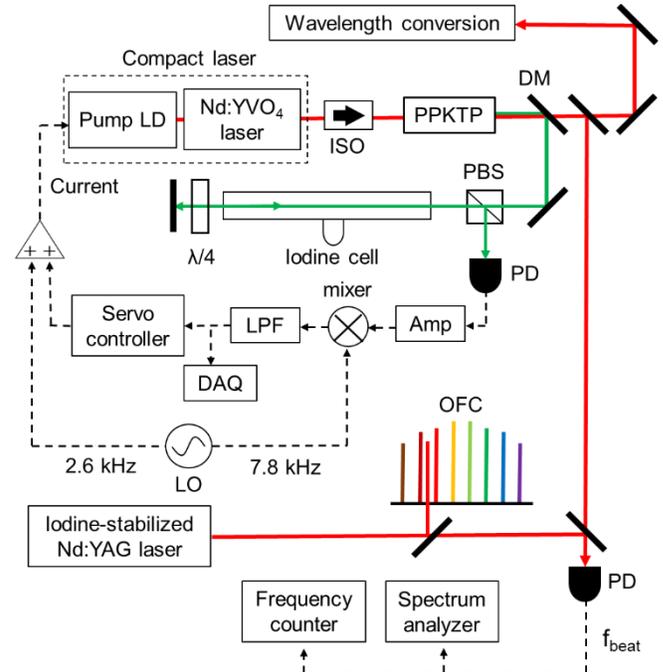

Fig. 2. Schematic diagram of a compact frequency-stabilized laser. Optical and electrical paths are shown as solid and dashed lines, respectively. ISO: Isolator, PPKTP: Periodically poled KTiOPO4, DM: Dichroic mirror, PBS: Polarizing beam splitter, PD: Photodetector, LO: Local oscillator, LPF: Low-pass filter, DAQ: Data-acquisition device, Amp: Amplifier, $\lambda/4$: Quarter wave plate, $f_{beat}$: beat frequency, LD: Laser diode, OFC: Optical frequency comb.

## 3. EXPERIMENTAL RESULTS

Figure 3 shows the observed third derivative saturation absorption signal of the $a_{10}$ hyperfine component of the R(86)33-0 transition of molecular iodine at 532 nm. The continuous frequency scan across the hyperfine component was obtained by tuning the injection current of the pump LD in the Nd:YVO4 laser. The optical power of the pump and probe beams was about 3 mW. The collimated beam diameters were ~ 1 mm. The cold-finger temperature of the iodine cell was held at 7 °C, corresponding to an iodine pressure of 8 Pa. The temperature of the cell body was matched to a controlled room temperature of 24 °C. For the even J number of the ground state (J = 86), the rovibrational energy level is split into 15 sublevels, resulting in 15 hyperfine components for the R(86)33-0 transition. The hyperfine component $a_{10}$ was chosen for laser frequency stabilization because it is located near the Doppler line center while being well isolated from the other hyperfine components. The signal-to-noise ratio of the $a_{10}$ component was approximately 78 in a bandwidth of 100 Hz. The linewidth of the $a_{10}$ component was ~ 1 MHz. We realized frequency stabilization by the feedback control of the injection current of the pump LD using the observed third derivative signal. Figure 4(a) shows the variation in the measured beat frequency between the compact laser and the iodine-stabilized Nd:YAG laser at 1064 nm during a one-hour long period, when the compact laser was free running or frequency locked. Fig. 4(b) shows the measured beat frequency for 12 h when the compact laser was frequency locked. The counter gate time was 1 s. Frequency variation of the beat frequency was observed over 38 MHz in the free-running case, while it was suppressed to 3 kHz (12 kHz) for the 1-h (12-h) long observation period in the frequency-locked case. Since the frequency stability of the iodine-stabilized Nd:YAG laser is better than $10^{-13}$ (corresponding to an uncertainty of the laser frequency of < 100 Hz) at an averaging time τ of 1 s [17], the frequency stability of the observed beat frequency is mainly dominated by the compact laser. Figure 4(c) shows the Allan standard deviation calculated from the measured beat frequencies as a function of the averaging time. The frequency stability of the free-running compact laser is $5.5 \times 10^{-10}$ at τ = 1 s (corresponding to an uncertainty of the laser frequency of 160 kHz), increasing over τ = 700 s towards $2.6 \times 10^{-8}$. The dashed line shows the frequency stability requirement ($3.5 \times 10^{-9}$) of the linewidth of the NV centers in diamond (< 1/10 of the linewidth). The frequency stability of the free-running compact laser no longer meets the requirement over τ = 30 s. However, the frequency stability of the frequency-locked compact laser is $6.5 \times 10^{-13}$ at τ = 1 s (corresponding to an uncertainty of the laser frequency of 180 Hz), achieving an improvement of about 3 orders of magnitude compared to that of the free-running case. The minimum and maximum Allan standard deviation of the frequency-locked compact laser are about $4.0 \times 10^{-13}$ (τ = 13 s) and $5.7 \times 10^{-12}$ (τ = 3600 s), respectively. The maximum Allan standard deviation of the frequency-locked compact laser is still more than two orders of magnitude smaller compared with the requirements of the NV centers in diamond. The improvement of the frequency stability of the frequency-locked compact laser over τ = 3600 s guarantees that the frequency stability of the compact laser meets the requirements of the NV centers for long-term operation.

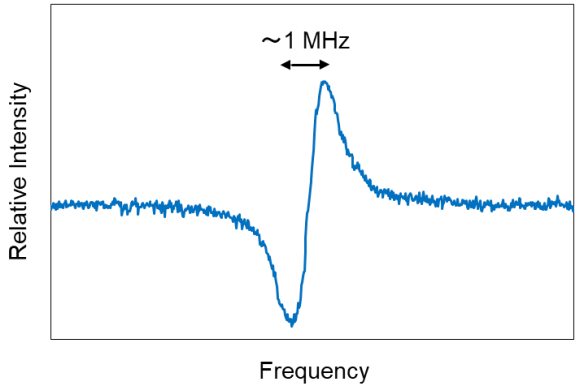

Fig. 3. Third derivative saturated absorption signal of the $a_{10}$ hyperfine component of the R(86)33-0 transition observed with an acquisition time constant of 10 ms. The linewidth and the signal-to-noise ratio are about 1 MHz and 78, respectively. The sweep time for the data acquisition was about 1 s.

The absolute frequency of the frequency-locked compact laser can be calculated from the measured beat frequency between the compact laser and the iodine-stabilized Nd:YAG laser. The frequency of the fundamental light of the iodine-stabilized Nd:YAG laser locked to the $a_{10}$ component of the R(86)33-0 transition was measured in advance to be 281,614,322,178.5 kHz ± 0.5 kHz by using an optical frequency-comb in our laboratory. We then calculated the frequency of the fundamental light of the compact laser locked to the same hyperfine component of the R(86)33-0 transition to be 281,614,302,191.1 kHz ± 1.6 kHz. The frequency uncertainty (relatively $5.7 \times 10^{-12}$) is mainly contributed from the maximum Allan standard deviation of the frequency-locked compact laser. The frequency difference (12.6 kHz) between the compact laser and the Nd:YAG laser locked to the same molecular transition is considered to be caused by the difference of the iodine cells and that of the spectroscopic methods used in the different iodine-stabilized lasers. Nevertheless, the frequency difference is much smaller compared with the frequency requirement of the NV centers in diamond. This guarantees the frequency reproducibility of the pump lasers used in separate quantum repeater nodes.

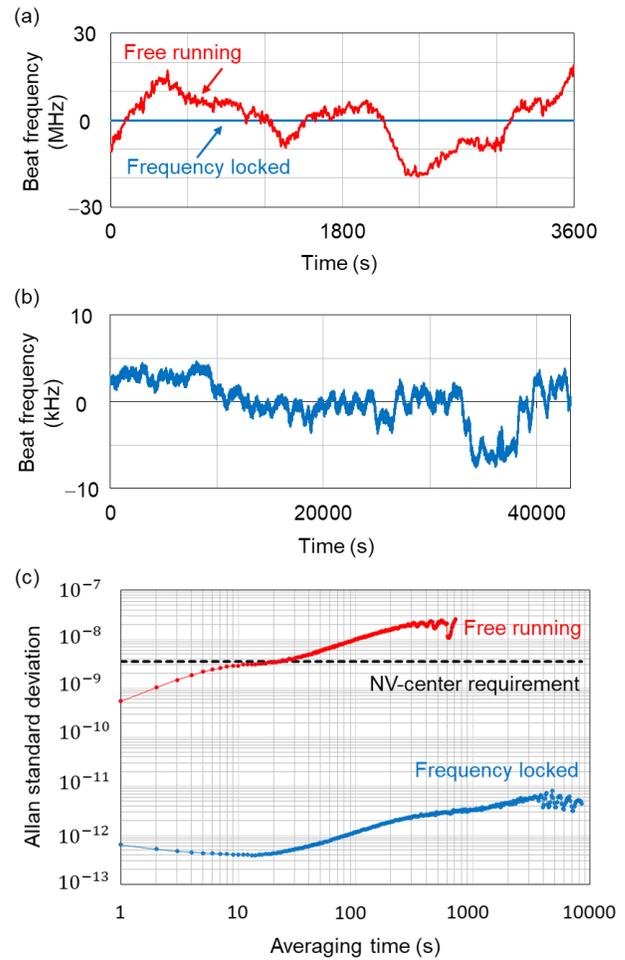

Fig. 4. Variations in the measured beat frequencies between the compact solid-state laser and iodine-stabilized Nd:YAG laser (a) when the compact laser was free running and frequency locked for 1 h. (b) When the compact laser was frequency locked for 12 h. Indicated data are the frequency deviation from the average. (c) Allan standard deviation calculated from the measured beat frequency between the compact and iodine-stabilized Nd:YAG lasers when the compact laser was free running (data in (a)) and frequency locked (data in (b)). The dashed line shows the NV centers requirements.

Figure 5 shows the spectrum of the observed beat frequency between the locked compact laser and the iodine-stabilized Nd:YAG laser. The resolution bandwidth of the spectrum analyzer was 3 kHz. Since the linewidth of the Nd:YAG laser is at kHz level, the observed beat linewidth of about 400 kHz is mainly contributed from the compact laser. The downward convex at the top of the beat indicates that the beat linewidth is caused by the frequency modulation applied to the compact laser. Again, the linewidth of the frequency-stabilized compact laser is much smaller than required for the NV centers. We note that, in the case of Ref. [15], the linewidth was about 1-2 MHz that is close or beyond the NV-center requirement.

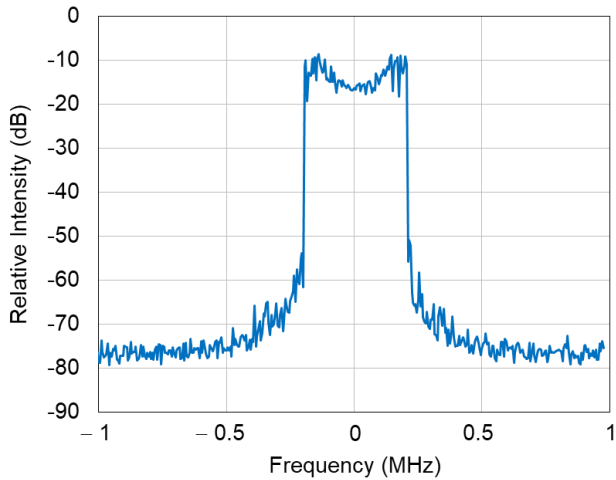

Fig. 5. Spectrum of the beat frequency between the locked compact laser and iodine-stabilized Nd:YAG laser. The resolution bandwidth was 3 kHz.

## 4. DISCUSSION AND CONCLUSION

As shown in Fig. 4(c), the frequency instability of the iodine-stabilized compact laser increased from $4.0 \times 10^{-13}$ to $5.7 \times 10^{-12}$ in the averaging time ranging from 13 to 3600 s. This is considered to be due to the residual amplitude modulation (RAM) during the frequency modulation spectroscopy [18]. When RAM signals enter the frequency mixer combined with frequency modulation signals in the demodulation process, they result in a baseline offset in the third derivative absorption signal. If the RAM signals are time dependent, the baseline offset will also vary over time. This will cause a frequency drift due to the variation of the target voltage in the servo system, and subsequently increase the frequency instability of the iodine-stabilized compact laser. To decrease the frequency instability caused by the RAM, it is important to reduce not only its time variation but also the RAM itself. In the present experiment, RAM is considered to be generated by two processes: i) etalon effects in the spectrometer, ii) laser intensity modulation induced by the current modulation of the pump LD. To reduce the etalon effects in the spectrometer, we have tried to tilt the surface of the optical elements relative to the laser beam inside the spectrometer. However, laser intensity modulation cannot be reduced because the current modulation applied on the pump LD is necessary to introduce frequency modulation for the spectroscopy. A related issue is the frequency drift of the laser caused by the variation of the laser power shift (AC Stark shift) originating from the servo. During the frequency stabilization, the feedback signal changes not only the laser frequency but also the laser power that determines the shift of the transition frequency of iodine depending on the laser power during the spectroscopy. These are the limitations of the current spectroscopic method. However, as pointed out in the previous section, for $\tau > 3600$ s, these effects start to be averaged and the laser instability decreases with an increasing $\tau$. Therefore, the present frequency stabilization technique is sufficient for connecting distant NV centers.

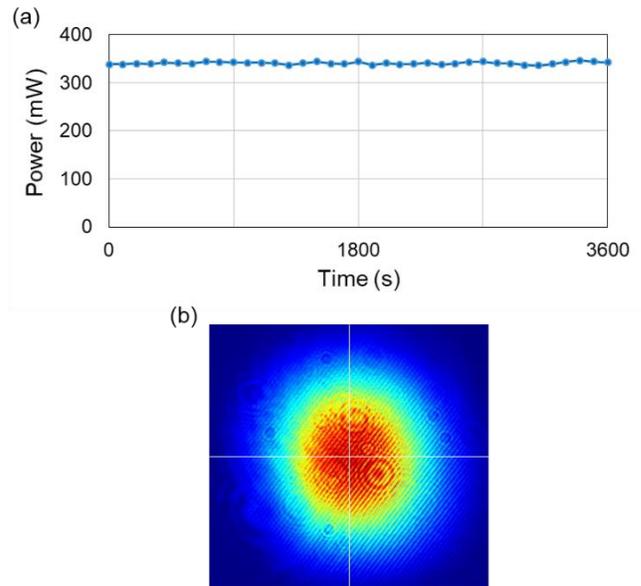

Fig. 6. (a) Long-term stability of the fundamental output power. (b) Transverse-mode profile of the fundamental beam.

Since the fundamental laser beam of the compact laser is used for the wavelength conversion for long-distance quantum communication, we evaluated the laser power and beam quality of the laser beam after travelling through a PPKTP crystal and several lossy optical components. Fig. 6(a) shows the laser power after encountering a dichroic mirror as a function of time. The optical power of the fundamental wavelength was over 300 mW and no degradation was observed over 1 h, which is promising for high-efficient and stable wavelength conversion. Fig. 6(b) shows the transverse mode profile of the laser beam measured with a charge-coupled device camera. The profile was well-fitted by a Gaussian function with an ellipticity of about 0.9. The observed beam profile is sufficient for obtaining a high-efficiency coupling to a wavelength conversion medium, such as a waveguide-type periodically poled lithium niobate (PPLN). In a previous work by several authors of the present study [19], the maximum wavelength conversion between 637.2 nm and a telecommunication wavelength was achieved with a pump power of about 500 mW where another laser with a slightly different wavelength (1071 nm) was used. However, we also demonstrated the same wavelength conversion process with the present laser and achieved a maximum conversion efficiency up to 35 % with a pump laser power of about 300 mW. The difference of the necessary pump powers could be due to the different spatial profiles which led to different coupling efficiency to the PPLN waveguide. We believe this is one of the advantages of the developed compact laser in this work. For example, in the case of Ref. [15], although the obtained long-term frequency stability was similar to that of the present work, the 1062-nm optical power (100 mW) is not sufficient for wavelength conversion without an optical amplifier. However, the installation of optical amplifiers in quantum repeater nodes would increase the cost of the equipment.

The present compact frequency-stabilized laser system fulfills the requirements necessary for being utilized in a quantum repeater node where an input telecom photon is converted into a visible photon to be absorbed by an NV center in a diamond, or vice versa. However, when two remote quantum memories need to be connected by a photon, the frequency detuning between the two memories becomes a problem. The degree of the detuning can reach several GHz. Therefore, usually the emitted photon from one NV center cannot be absorbed by another one. This detuning problem can be overcome by using pump lasers. Wavelength conversion is a process of sum and difference frequency generation between the signal from an NV center and a pump laser. Therefore, the detuning between the NV centers can be compensated by the detuning of the pump lasers. Before the entire procedure, two quantum memories' emitting/absorbing frequencies are determined by wavemeters with an accuracy around 100 MHz. An AOM modulates the pump laser frequency at the remote node to maximize the absorption of the photon obtained by sum frequency generation between the remote node pump photon and the telecom photon from the emitting node. If the absolute frequencies of the pump lasers are determined, the detuning frequency of the NV center at the remote node is precisely determined. When the detuning is too large to be compensated by a pump laser at the remote node, whose frequency is stabilized using the same absorption line as the pump laser at the emitting node, we can select a different hyperfine component or another transition resulting in a smaller detuning. The rich spectrum of molecular iodine containing a lot of absorption lines helps to obtain a detuning smaller than hundreds of MHz. Residual frequency discrepancy can be further compensated by using an AOM.

In long-distance quantum communication networks, a large number of quantum repeater nodes are required, since the distance between each node should be in the order of km [20]. Therefore, compactness and low cost of the wavelength conversion system—including a frequency-stabilized laser system inside each repeater node—are crucial for quantum repeater systems.

In conclusion, we have demonstrated an iodine-stabilized pump laser for wavelength conversion connecting NV centers in diamonds by photons in the telecom band. In addition, the obtained frequency-stabilized 1064-nm light shows a good transverse mode profile and sufficient power for wavelength conversion. This compact, low-cost and high-power frequency-stabilized laser is suited to be installed in quantum repeater nodes.

**Funding sources and acknowledgments.** Formal funding sources should be listed in a separate paragraph block before any other acknowledgment information. Funding sources and any associated grant numbers should match the information entered into the Prism **Funding Information.**

This work was supported by Toray Science foundation, the Asahi Glass foundation, KDDI foundation, the Murata Science foundation, JKA, REFEC and the Japan Society for the Promotion of Science (JSPS) KAKENHI (15H02028).

**Acknowledgment**. We thank Y. Yamamoto, S. Utsunomiya, T. Kobayashi, M. Fraser, and I. Iwakura for their support about the use of equipment.